\documentclass[final,3p,times]{elsarticle}
\usepackage[pdftex,
			colorlinks={true},
			pdfstartview={FitH}
            ]{hyperref}

\makeatletter
\def\ps@pprintTitle{%
  \let\@oddhead\@empty
  \let\@evenhead\@empty
  \def\@oddfoot{\reset@font\hfil\thepage\hfil}
  \let\@evenfoot\@oddfoot
}
\makeatother

\usepackage{graphicx}

\usepackage{paralist}
\usepackage{longtable}
\usepackage{array}
\usepackage{epstopdf}

\usepackage{amssymb}

\begin{document}
\begin{frontmatter}



\title{Pouring Cloud Virtualization Security Inside Out}


\author{Yasir Shoaib}
\ead{yasir.shoaib@ryerson.ca}
\author{Olivia Das}
\ead{odas@ee.ryerson.ca}
\address{Department of Electrical and Computer Engineering, Ryerson University, Toronto, ON M5B 2K3}

\begin{abstract}
In this article, virtualization security concerns in the cloud computing domain are reviewed. The focus is toward virtual machine (VM) security where attacks and vulnerabilities such as VM escape, VM hopping, cross-VM side-channel, VM-based rootkits (VMBRs), VM mobility, and VM remote are mentioned and discussed according to their relevance in the clouds. For each attack we outline how they affect the security of cloud systems. Countermeasures and security measures to detect or prevent them through techniques such as VM detection, GuardHype, VM introspection, VM image scanning, etc. are also discussed. Through the surveyed work we present a classification of VM threats within the clouds. Finally, we include our observations and those of other researchers on this matter of cloud virtualization security.
\end{abstract}


\end{frontmatter}














\section{Introduction}
\label{sec:Introduction}
Many up to now would have heard about cloud computing and those with a keen sense of curiosity and belief in this new paradigm's capabilities would have experienced it and embarked on a journey of --- dilemma at first, followed by --- excitement and hopeful possibilities. Others may have been skeptical, posing a question like this one: ``how can a mishmash of existing technologies just be hailed as a new technology?'' For those with similar reservations, cloud computing may just be a ``buzz'' \cite{2009_CloutCloud,2009_CloudNewWineNewBottle} surrounded by confusion with relating ideas \cite{2010_SecuringElasticityCloud,2010_OverviewVirtualCloudComputing}. Reasons to believe that the ``cloud'' \cite{zhang2010cloud13} is nothing new, lies in its similarities to paradigms such as grid computing \cite{2008_GridComputingSecurityTaxonomy} and to some extent with the cloud computing idea itself. John McCarthy \cite{2012_JohnMcCarthy}, recipient of 1972 Turing award and computer scientist who coined the term ``aritifical intelligence,'' in a 1961 speech foresighted a possible form of computing, as a future of time-sharing, that may come to be delivered as a ``utility'' \cite{2012_JohnMcCarthy,2009_InCloudsShallWeTrust,2010_SecurityInTheCloud,zhang2010cloud13}. Cloud computing  brings to life such an idea. But, there is more to it than just time-sharing and utility based computing, and the impact due to the mere existence of such a paradigm is nothing less than far-reaching. Those who have realized this and understood what clouds really are have relinquished their initial doubts and have come to embrace cloud computing as believers \cite{2010_SecuringElasticityCloud}.

For application providers (AP) developing and deploying web applications, the cloud paradigm facilitates the use of computing resources from public cloud providers (CP) with ``multi-dimensional ease'' \cite{2012_ucc2012YasDas}. With the ability to allow for quick resource addition and removal, a feature which is known as elasticity \cite{NIST_CloudDefn}, the cloud systems provide a unique capability to their customers \cite{2012_MeasureElasticity,2010_SecuringElasticityCloud}. Furthermore, with various service and deployment models to choose from, such as those models mentioned in ``The NIST definition of cloud computing'' \cite{NIST_CloudDefn}, a customer can either hand-pick desired services from service providers or themselves setup a private cloud infrastructure for their own use.

\begin{figure*}[h!tb]
\centering
\includegraphics[keepaspectratio,width=\textwidth]{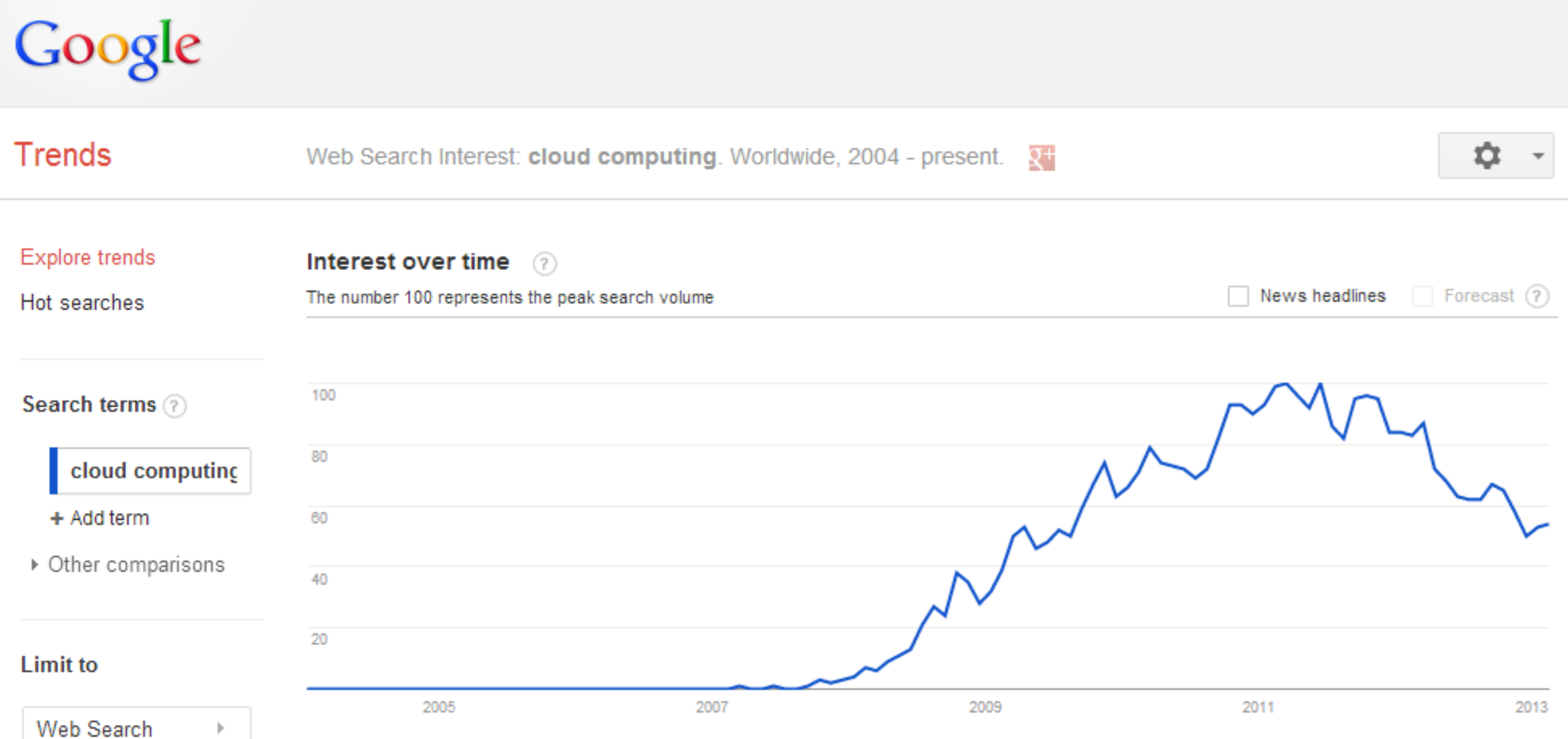}
\caption{Trends of search term ``cloud computing'' by users worldwide on Google Web Search. For insufficient number of searches, data may not be visible on the graph, which may be the case for searches before year 2007. Provided by Google Trends \cite{2013_GoogleTrends}. Retrieved on Feb 12, 2013. \copyright 2013 Google. Google and the Google logo are registered trademarks of Google Inc., used with permission.}
\label{fig:CC_Trends_ALL}
\end{figure*}

Over the past few years, interest in cloud computing has soared, mainly due to points highlighted above. \figurename~\ref{fig:CC_Trends_ALL} shows a graph obtained from Google Trends \cite{2013_GoogleTrends} on the term ``cloud computing'', indicating increase in Google web searches performed by users worldwide on the topic. Year 2007 marks the beginning of significant search requests made to Google so as to be seen on the graph, although, cloud computing had come to be known by many since 2006 \cite{zhang2010cloud13}. Peak searches have been performed during the first quarter of 2011, although, not as many searches have been launched using the particular keyword since then.

With growing popularity, focus has also moved toward the challenges that are being faced by cloud computing. Security is one domain that has received a lot of concerns \cite{2009_CloudSecurityInfoWeek,2012_SecurityChallengesPublicCloud}. From Google Trends \cite{2013_GoogleTrends} it has been found that in the past 12 months, searches for terms: ``cloud security'' and ``cloud computing security'' have risen by +50\% and +40\%, respectively. Many raise cloud security issues while visualizing the concerns from different perspectives such as trust \cite{2009_InCloudsShallWeTrust}, authorization, multi-tenancy, access control \cite{2010_TowarsMultiTenancyAuthCloud}, virtualization \cite{2012_ThreatAsAServiceVirtImpact}, etc. In this article, we focus toward virtualization security relating to virtual machines (VMs) in the clouds. From previous publications, we consider attacks and vulnerabilities encountered in virtual environments and describe how these attacks relate to cloud systems, alongside discussing specific cloud virtualization threats. Furthermore, how those security issues can be dealt with are discussed. A classification based on the surveyed works on VM threats within the clouds has also been presented.

This article is organized as follows: Section~\ref{sec:CloudComputing} describes cloud computing technologies. Section~\ref{sec:Virtualization} introduces virtualization. Section~\ref{sec:VirtualizationSecurity} reviews virtualization security in the clouds. In Section~\ref{sec:ClassificationVMThreats}, we classify the threats on VMs in the clouds.  Section~\ref{sec:SummaryRelatedWorks} provides a summary of the works surveyed in this article. In Section~\ref{sec:ObsCloudSecurity}, we include our observations and those of security researchers in the area of cloud virtualization security. Finally, Section~\ref{sec:Conclusions} presents the conclusions.

\section{Cloud computing}
\label{sec:CloudComputing}
Cloud computing is based on technologies such as virtualization and requires a network, viz. the Internet, to provide one or more of following services: IaaS (Infrastructure-as-a-Service), PaaS (Platform-as-a-Service) and SaaS (Software-as-a-Service) \cite{zhang2010cloud13}. IaaS allows customers access to CPU, storage resources and networking via virtualization of physical hardware, in the form of VMs. VMs may be quickly allocated (and de-allocated) as needed by the customers; this is possible through the features of elasticity and ``[o]n-demand self-service'' \cite{NIST_CloudDefn}. PaaS provides customers with a functioning environment --- consisting of operating system (OS) and software development kit (SDK) --- for development and deployment of software services \cite{zhang2010cloud13}. SaaS provide end-users with access to web based software that run on the cloud data centers. Virtualization and databases help in providing multi-tenant support, i.e. multiple users may have access to shared resources \cite{2012_EnforcingMultitenancyCloud}.

Clouds may be deployed as a public-cloud, community-cloud, private-cloud or hybrid cloud \cite{NIST_CloudDefn}. Public-clouds are provided by organizations such as Amazon, Google and Microsoft who host hardware and software resources in their facilities; although the levels of services offered may differ from one provider to another \cite{ranjan2010peer12,zhang2010cloud13}. Customers pay CP as per the amount of resources used, an alternative to paying for capital expenses  \cite{velte2009cloud1,2012_ucc2012YasDas}. Private clouds are hosted by individual organizations to serve their own internal users, whereas community-clouds are hosted for users in a group of organizations \cite{NIST_CloudDefn}. Hybrid clouds are composed as a combination of previously mentioned deployment models \cite{NIST_CloudDefn}.

\section{Virtualization}
\label{sec:Virtualization}
Virtualization technology offers sharing of hardware for running of isolated guest OSs. A virtual machine monitor (VMM), a.k.a. hypervisor, is positioned  between the host hardware and one or more guest OSs, and manages access of shared resources needed by the guests \cite{2005_ResourceVirtRenaissance,2005_IntelVT,2005_VirtualMachineMonitorsCurrentTechFutureTrends}. Each OS runs as a separate VM instance, which is an encapsulated platform running on top of a hypervisor. One or more OSs execute under the control of the hypervisor, where each OS is enclosed within its own VM, interacting with the hardware indirectly through the hypervisor \cite{2010_InVMMeasuringFramework,2008_MeetTheVirts}. When the hypervisor executes right above the physical hardware, the arrangement is known as Type-I virtual environment \cite{2008_ParadoxOfSecurityInVM}. If the hypervisor executes within a host OS then such an environment is known as Type-II \cite{2008_ParadoxOfSecurityInVM}.

Virtualization de-couples the physical machine from the VM and as a consequence the latter can be live-migrated to another machine, copied as a VM image on disk to another VM image, taken snapshot of for restoration purposes later, etc. altogether making possible various operations that allow for enhancement of non-functional attributes, e.g. performance, fault-tolerance. 

Cloud computing is build on top of virtualization, which is a ``core'' \cite{2012_ThreatAsAServiceVirtImpact} technology of the cloud. Multiple VMs running on shared physical resources such as CPUs, memory, networks etc. are isolated from each other due to virtualization, thereby providing security. Key functionalities made possible through virtualization are workload consolidation, workload isolation and workload migration \cite{2005_IntelVT}. These form an inherent part of the cloud operations for delivering security, resource provisioning, power management and performance.

In cloud computing, for launching VM instances, customers (generally) use a bank of VM disk images made available by their cloud platforms \cite{2010_Sempolinski_ComparisonOpenEucNim}; however, options may be available to create and upload custom developed VM images to the cloud, which depends on the cloud platform being used and the support offered for custom images. An example of publicly shared VM images is Amazon Machine Images (AMI) \cite{2013_AmazonAMI} for the Amazon EC2 cloud \cite{2013_AmazonEC2}, where images are uploaded by organizations and users (a.k.a. ``Publishers'' \cite{2011_Bugiel_VMimages}). In the former, an image can be created/published from a running VM instance by taking its snapshot \cite{2011_Bugiel_VMimages}. Customers have various options to choose from in the images database to obtain their desired software stack.

\section{Virtualization security in clouds}
\label{sec:VirtualizationSecurity}
Virtualization security has been a topic of discussion for a while, usually in its own context, i.e. without linking it to cloud computing, considering the fact that virtualization existed before being one of backbones of cloud computing (e.g. \cite{2008_TamingVirtualization,2008_ParadoxOfSecurityInVM,2008_VirtSparksSecurityConcerns,2007_HidingVirtualizationFromAttackersMalware}). More recently though, since the onset of cloud computing, researchers have been coupling the two in their investigation and findings (e.g. \cite{2012_ThreatAsAServiceVirtImpact,2009_HeyYouGetOffCloud,2010_SecurityInMultiTenancyCloud}). Here, we discuss the security related topics from published works. Of the works and the attacks surveyed in this article, the following are the most cloud-specific threats: ``VM hopping'' \cite{2010_SecurityInMultiTenancyCloud,2012_ThreatAsAServiceVirtImpact}, ``cross-VM side-channel attacks''\cite{2009_HeyYouGetOffCloud}, ``VM Mobility'' \cite{2012_ThreatAsAServiceVirtImpact}, and other VM image attacks \cite{2011_Bugiel_VMimages}. Although, not specific to only cloud systems, ``Blue Pill''  \cite{2006_IntrodBluePill,2006_SuvertingVistaJR} VMBR does pose threat to cloud systems, and has been listed as an example of ``Shared Technology Issues'' \cite{2010_CSATopThreats} threat by Cloud Security Alliance (CSA). 

We begin our discussions with general virtualization security concerns (e.g. VM detection) and then systematically move to various distinctive cloud related threats, those which were listed in the above paragraph.

\subsection{VM detection}
\label{subsec:VMDetection}
Virtual machines provide isolation, both in terms of inter-VM and host-to-VM isolation. This feature has really served the purpose of security researchers, who instead of running unknown malware directly on bare hardware OS, would run it in a VM \cite{2007_HidingVirtualizationFromAttackersMalware}. This allows for a easy analysis of the malware, in a rather constrained environment, with the ability to repeat the process for an even thorough analysis with reinstated VM image. However, Carpenter et al. \cite{2007_HidingVirtualizationFromAttackersMalware} point out that malwares are now using VM detection intelligence and they choose their actions/behaviour accordingly. One course of action by the malwares is hiding their disruptive functionality within VM environments, thereby avoiding themselves from detection or analysis \cite{2007_HidingVirtualizationFromAttackersMalware}. Carpenter et al. \cite{2007_HidingVirtualizationFromAttackersMalware} mention use of host-VM communication channel and Red Pill \cite{2007_VirtualisationAsBlackhatTool} (introduced by Joanna Rutkowska \cite{2012_InvisibleThingsLab,2012_InvisibleBlogSpot}) as two approaches for VM detection running in VMware.

\subsection{VM escape and VM hopping in clouds}
\label{subsec:VMescapeVMhopping}
Although, malwares could detect VMs and remain dormant, there are other options for the more intelligent and dangerous malwares: attacking vulnerabilities in the VM environment \cite{2007_HidingVirtualizationFromAttackersMalware}. In a cloud domain, where VMs are ubiquitous, VM detection does not have to precede an attack. ``VM escape'' attack \cite{2010_SecurityInMultiTenancyCloud,2010_VirtualizationSurveyConceptsTaxonomy,2007_HidingVirtualizationFromAttackersMalware} would allow an attacker's malware to escape the VM to the host or hypervisor on which the VM is running by exploiting vulnerabilities in the environment. A variant would be ``VM hopping'' \cite{2010_SecurityInMultiTenancyCloud,2012_ThreatAsAServiceVirtImpact} allowing hopping of the malware's attacker from a VM to another peer VM co-resident on the same host or within the control of a common hypervisor. Furthermore, VM hopping attacks the ``confidentiality, integrity and availability'' \cite{2012_ThreatAsAServiceVirtImpact} of  the VMs, where the attacker has vast control options once they are inside a VM. In the context of cloud computing, these attacks can have serious repercussions where all aspects of security need enforcement.

\subsection{VM detection avoidance}
\label{subsec:VMdetectionAvoidance}
While detection of a VM environment (or ``VME'' \cite{2007_HidingVirtualizationFromAttackersMalware}) from within a VM would benefit attackers, approaches to circumvent detection have also been proposed. In this context, Carpenter et al. \cite{2007_HidingVirtualizationFromAttackersMalware} use two approaches in VMware environment by using ``VMware's undocumented features and modifying the VMware binary program \ldots'' \cite{2007_HidingVirtualizationFromAttackersMalware}. In the former option, configuration files (VMX files) which are under the control of the VM administrator, are modified to make changes to memory relocation and binary translation functions to achieve the VM hiding objective. The latter method is achieved by modifying a parameter value (specifically the ``VMXh'' value \cite{2007_HidingVirtualizationFromAttackersMalware}) in the VMware binary and VM image, to the value chosen by the user which the attacker cannot then use to exploit the host-VM communication channel vulnerability. The authors use ``VMmutate''\cite{2007_HidingVirtualizationFromAttackersMalware}, a tool they have developed to achieve binary translation. These efforts, however, lead to decrease in the following features: ``drag-and-drop, cut-and-paste via the clipboard, and shared file directories'' \cite{2007_HidingVirtualizationFromAttackersMalware}.

\subsection{VM rootkits and the clouds}
\label{subsec:VMrootkits}
Imagine a malware that runs undetected, lurking for a golden opportunity to spring up or just remains invisible but watching every possible move. Ford \& Allen \cite{2007_HowNotToBeSeenI,2007_HowNotToBeSeen-II} in their two part article discuss about this notion of ``stealth'' \cite{2007_HowNotToBeSeenI,2007_HowNotToBeSeen-II}. They begin with some beneficial uses of stealth (e.g. hiding of application by vendors or for security purposes), however, they come toward the main issue about its use by attackers. They eventually bring the discussion toward a new class of malware: VM-based rootkits (``VMBR'' \cite{2006_SubVirt}). The idea behind these VM-based rootkits is taking control of an operating system running on bare hardware, by having it execute under the attacker's hypervisor, thereby maintaining their stealthiness \cite{2006_SubVirt,2006_SuvertingVistaJR,2006_Vitriol,2007_HowNotToBeSeenI,2007_HowNotToBeSeen-II}. Such VM rootkits  became popular in the year 2006 because of these following techniques: SubVirt \cite{2006_SubVirt}, Blue Pill \cite{2006_IntrodBluePill,2006_SuvertingVistaJR}, and Vitriol \cite{2006_Vitriol}. SubVirt was used in controlling of Windows XP and Linux OS through VirtualPC and VMware VMBRs, respectively, on x86 platform \cite{2006_SubVirt}. Blue pill demonstrated to ``subvert'' \cite{2006_SubVirt} the Windows Vista OS on AMD64 \cite{2006_SuvertingVistaJR}. Vitriol targeted MacOS X on Intel Core Duo/Solo \cite{2006_Vitriol}. In the following paragraphs, we focus our discussions on Blue Pill.



Blue Pill was introduced to the world by Joanna Rutkowska through her blog \cite{2006_IntrodBluePill}. Interested readers may refer to the following \cite{2006_IntrodBluePill,2006_BlogLog} to read what she had to say about Blue Pill on the release. The novelty of Blue Pill in comparison to the earlier VMBR, i.e. SubVirt, is that it installs without a restart and without modifying the contents of the storage disk, thereby challenging detection \cite{2007_VirtualisationAsBlackhatTool,2006_SuvertingVistaJR}. However, Blue Pill would fail if the system was restarted \cite{2006_SuvertingVistaJR}. To circumvent these problems, Rutkowska indicates that the shutdown can be intercepted by the hypervisor to emulate a shutdown sequence of the OS, and that the hypervisor itself survives the reboot \cite{2006_SuvertingVistaJR}. An issue which Rutkowska mentions is the problem with nested virtualization, where an OS was already running on an hypervisor and if Blue Pill was tried on it, then this would possibly crash the system (if nested virtualization was not supported); therefore Blue Pill had to support nested virtualization to avoid detection by crash \cite{2006_SuvertingVistaJR}. Well, in Rutkowska's March 31, 2008 blog post \cite{2008_KickHypervisor}, that nested hypervisor issue became a thing of the past. The announcement was support of nested virtualization by Blue Pill, which was complemented by a snapshot of Windows XP running inside Windows Vista (which was itself ``bluepilled'') \cite{2008_KickHypervisor}.

The concept behind Blue Pill is nothing but novel, however, question is raised about it being a possible threat to cloud computing. For this, consider a malicious action by an attacker, who installs Blue Pill on a purchased IaaS VM to oust control from the cloud's hypervisor, or more correctly, run through nested virtualization, and take control of not only their own VM, but also co-resident VMs.  Bradbury \cite{2010_VirtualAttacks} points out with regards to Blue Pill that the ``attack could also be used to compromise an existing hypervisor'' \cite{2010_VirtualAttacks}, suggesting such actions are possible. Furthermore, Cloud Security Alliance (CSA), which published a list of top seven cloud computing threats, listed Blue Pill as an example of ``Shared Technology Issues'' threat \cite{2010_CSATopThreats}. Note, Blue Pill does require some specialized hardware features such as AMD's SVM functionality to work, but support for Intel VT-x processors has also been added \cite{2008_Vegas2008}. With the nested hypervisor working, however, unknown threats could materialize in the future. For this, we will have to wait for further developments in the Blue Pill concept.

\subsection{VM detection -- revisited}
\label{subsec:VMdetectionRevisited}
While VMBRs remain a threat, their prevention and detection are key to safeguarding systems. One general approach is to determine if the execution is happening within a VM, where it is expected not to be. Interestingly, Red Pill developed by Joanna Rutkowska (refer Section~\ref{subsec:VMDetection}) itself can spot SubVirt \cite{2007_VirtualisationAsBlackhatTool}. This is possible via the SIDT instruction (or SxDT instructions \cite{2006_SuvertingVistaJR,2005_IntelVT}), which returns a ``unique'' \cite{2007_VirtualisationAsBlackhatTool} Interrupt Descriptor Table Register (IDTR) address when running within a VM, different from when it is executed natively \cite{2007_VirtualisationAsBlackhatTool,2007_HowNotToBeSeen-II,2005_IntelVT}. The reason is that SubVirt targets x86 architecture, which does not run in full virtualization and in such case reading from privileged registers is possible from an unprivileged level, i.e. from a software within a guest VM, thereby helping with virtual environment detection \cite{2007_VirtualisationAsBlackhatTool,2005_IntelVT}. This technique, however, cannot be applied for detection of Blue Pill, which runs in full virtualization mode \cite{2006_SuvertingVistaJR}. In fact, this was the goal in development of Blue Pill, to hide, to be stealthy \cite{2006_SuvertingVistaJR}. Rutkowska listed two levels of stealth \cite{2006_SuvertingVistaJR}:
\begin{description}
\item[Level 1 (BPL1)] Hiding when idea is known.
\item[Level 2 (BPL2)] Hiding when source code is available. BPL2 was not implemented then.
\end{description}

Focusing on BPL1, there are still some interesting detection options. Allen \& Ford \cite{2007_HowNotToBeSeen-II} refer to work done by Garfinkel et al. \cite{2007_CompatibilityIsNotTransparency} in this regard. Monitoring of Translation-Lookaside Buffer (TLB) hits and misses and timing variance within VMs are good candidates \cite{2007_HowNotToBeSeen-II}. Running in a VM will cause timing affects compared to a native system, giving away existence of virtualization. With regards to timing, Rutkowska based on this has suggested earlier of employing timing dilation implementation to avoid detection \cite{2006_SuvertingVistaJR}. However, Garfinkel et al. \cite{2007_CompatibilityIsNotTransparency} argue that time dilation has overhead with affects on performance and gives higher chances of the malicious hypervisor's detection. Allen \& Ford \cite{2007_HowNotToBeSeen-II}, however, bring up an important point that the mere installation of a malicious code in the kernel is a problem and if the kernel avoids installation of unsigned driver codes then VMBR related problems would be avoided.

For cloud computing, detection of VM to determine existence of VMBR would not work because VM is ubiquitous in the cloud. Carbone et al. \cite{2008_TamingVirtualization} suggest ``GuardHype''\cite{2008_TamingVirtualization} as an idea, which would allow only legitimate hypervisors and prevent malicious VMBRs. GuardHype is based on nested virtualization, stationing itself beneath any running hypervisior and controlling the activities, thereby guarding malicious activity by a hypervisor. Further research in this area is needed as a solution for preventing and spotting VMBRs like Blue Pill.

\subsection{VM images in clouds}
\label{subsec:VMimgesClouds}
Security concerns due to unintentional exposition of vulnerabilities through the use and publishing of publicly shared VM images in the clouds by na\"{\i}ve users (or customers) have been discussed and analyzed by Bugiel et al. \cite{2011_Bugiel_VMimages}. The authors use their custom developed software to analyze images and retrieve sensitive information present within. Although, their analysis has been conducted on Amazon EC2 cloud \cite{2013_AmazonEC2}, they mention that ``the methods and techniques described in [their] paper are applicable to arbitrary IaaS providers \ldots'' \cite{2011_Bugiel_VMimages}. From an attackers cost-benefit viewpoint, they also compute the small cost of their attacks in comparison to the quite large monetary loss that could potentially incurred by organizations because of the attacks. In their paper, mistakes such as losses of API keys, private keys (e.g. SSH and SSL), sensitive personal information and software source codes by means of unintentionally including them within public VM images and thus the associated information leakage threats, have been mentioned alongside with presentation of their analysis results from using their tool. Examples of threats include those emanating from malicious use of user SSH private keys to login to machines that allow access to those user credentials, or use of API keys to launch VM instances of the key owner and deploy distributed denial of service (DDoS) attacks on other machines. Alongside discussing the threats from sharing of user private keys, the authors also explain about other SSH specific threats and attacks, such as those due to inclusion of user public keys in images that then allow for SSH backdoors and inclusion of stale SSH host keys that lead to impersonation, man-in-the-middle attacks, etc. Furthermore, they do not restrict their discussions to unintentional causation behind creation of VM images, but also mention about malicious intentions behind creation of VM images by citing various related works. They mention that some actions, such as creation of SSH backdoors, inclusion of malware and introduction of vulnerabilities within the VM images (referred as ``[m]alicious VMs'' \cite{2011_Bugiel_VMimages}) may also be caused by malicious publishers.

While highlighting VM image security issues, Bugiel et al. \cite{2011_Bugiel_VMimages} have also suggested countermeasures against the attacks. Through ``awareness'' \cite{2011_Bugiel_VMimages} of the security issues that are aligned with handling of VM images, they intend to educate the customers about the involved risks. Also to help the process, they examine reasons behind inadvertently including of the keys within images (e.g. public, private keys and API keys). General countermeasure suggestions include scanning of VM images (i.e. ``VM image scanning'') by cloud providers for vulnerabilities, display of warning to publishers about inclusion of private keys, and rating of VM images according to their ``usefulness and quality'' \cite{2011_Bugiel_VMimages}.

From a cloud standpoint, these authors have highlighted the large-scale impact of security issues associated with vulnerable and malicious VM images; these public VM images are used by many customers, who may directly use a copy of the VM image to create VM instances or who may rely on a modified derivate of the image, where many images and instances may exist from a parent vulnerable/malicious VM image. Furthermore, threats may also emerge from within vulnerable or malicious live VM instances which are used to create VM images from their snapshots.

To add to the above discussion, Tsai et al. \cite{2012_ThreatAsAServiceVirtImpact} dub the cloning and movement of VM images as ``VM Mobility'' \cite{2012_ThreatAsAServiceVirtImpact}, where they mention the related attacks and vulnerabilities, and furthermore discuss about the impact on IaaS and other layers.

\subsection{Cross-VM side-channel attacks in clouds}
\label{subsec:VMsidechannel}
Ristenpart et al. \cite{2009_HeyYouGetOffCloud} present a compelling work highlighting vulnerabilities in VM environments due to multitenancy on physical machines in public clouds, and building on the information to introduce ``cross-VM side-channel attacks'' \cite{2009_HeyYouGetOffCloud}. The authors orchestrate the attacks on Amazon EC2 cloud \cite{2013_AmazonEC2}; however, they ``believe that variants of [their] techniques are likely to generalize to other services \ldots'' \cite{2009_HeyYouGetOffCloud}. The mentioned attacks include ``cross-VM keystroke monitoring'' \cite{2009_HeyYouGetOffCloud} and provide means of gathering confidential information. Few suggestions on how to prevent them are provided.

\subsection{VM remote attacks and clouds}
\label{subsec:VMremote}
Before exploiting vulnerabilities within a cloud environment and launching attacks across the cloud, an attacker needs to gain access to a VM running within the cloud system; although at times, the eventual target may just be the VM. In either circumstance, the attacker may use well-known remote attacks that work for any host/machine (VM or otherwise), and McClure et al. \cite{2012_mcclure_hackingExposed7} explain various ways that attackers may do just that. For remote attacks, a ``[r]emote access'' \cite[p.~234]{2012_mcclure_hackingExposed7} to the target system is needed through a network. For Amazon EC2 clouds, VMs have access to external network (Internet) and internal network (for inter-VM communication) through external and internal assigned IP addresses, respectively \cite{2011_Bugiel_VMimages,2009_HeyYouGetOffCloud}; other cloud system would have similar connectivity options for VMs such that they can be accessed and configured by the customers.

For remote attacks, one approach is finding network services that are executing on the system by scanning the network ports through easily available tools (e.g. Nmap) and then exploiting well-known vulnerabilities of that particular version of the service to enter the system \cite[pp.~1--6]{2012_mcclure_hackingExposed7}. After remote access, the next steps by the attacker may be toward getting root user access and carrying out further attacks from the compromised system \cite[p.~234]{2012_mcclure_hackingExposed7}. Ristenpart et al. with regards to ``direct compromise'' \cite{2009_HeyYouGetOffCloud} from remote attacks, acknowledge that such a ``threat exists for cloud applications as well'' \cite{2009_HeyYouGetOffCloud}. 

On a similar and related topic to remote attacks, Provos et al. \cite{2009_CyberCrimeWhenCloudTurnsDark} focus their attention on web attacks and cyber-crimes. The shift in motivation of attackers toward monetary benefits and the attacks on web-browsers and web servers are discussed. Discussions include: ``drive-by download'' \cite{2009_CyberCrimeWhenCloudTurnsDark}, SQL injection attacks, .htaccess redirection attacks, social engineering attacks, and challenges. Although the aforementioned article doesn't mention about cloud computing, the issues highlighted are quite relevant in the cloud domain and should not be ignored. To add, McClure et al. \cite[chap.~10]{2012_mcclure_hackingExposed7} also discuss about web and related attacks along with advice to safeguard against them.

Since clouds host SaaS applications, compromise of those services and eventual gain of access inside the cloud systems is a real threat. Furthermore, attackers who have once entered the system may employ any of the attacks suggested above: VM escape, VM hopping, cross-VM side-channel and VMBRs, to wreck havoc in the cloud domain.

\subsection{VM introspection and clouds}
\label{subsec:VMintrospection}
VM introspection is a way to keep an eye on activities relating to a VM \cite{2008_VirtualMachineIntrospectionObsInterference}. The introspection roles may include either simple monitoring and reporting of threats or performing actions toward mitigating those threats \cite{2008_VirtualMachineIntrospectionObsInterference}. Other aspects that distinguish VM introspection techniques from one another is extent of knowledge of OS running on the VM (a.k.a. ``semantic awareness'' \cite{2008_VirtualMachineIntrospectionObsInterference}) and replay of events for forensic analysis after a intrusion \cite{2008_VirtualMachineIntrospectionObsInterference}. These techniques are useful in cloud environments to analyze any attempts of an attack to the system, piece together the events that lead to such an attack, or to trace the steps of a compromise after the fact. Anthes \cite{2010_SecurityInTheCloud} discusses about benefits of such techniques in the cloud, where IBM researcher Matthias Schunter points out that instead of having many virus scanners running on each VM on a physical machine, only one ``protected VM'' \cite{2010_SecurityInTheCloud} could do the introspection job.

Christodorescu et al. \cite{2009_CloudSecurityIsNotJustVirtSecurity} introduces their approach of VM introspection that requires no prior knowledge of the guest OS running on the VMs or their security state, which was lacking in the previous introspection techniques. Their approach is useful specially in dynamic environments such as cloud computing where VM migration is common and security of the VM has to be evaluated, and managed. The authors demonstrate two applications of their introspection technique:
\begin{inparaenum}[i)]
\item VM OS identification, and
\item Rootkit detection and recovery.
\end{inparaenum} 
These are promising techniques monitoring and maintaining cloud security in the future.

\section{Classification: VM threats in clouds}
\label{sec:ClassificationVMThreats}
\begin{figure*}[!htb]
\centering
\includegraphics[keepaspectratio,width=\textwidth]{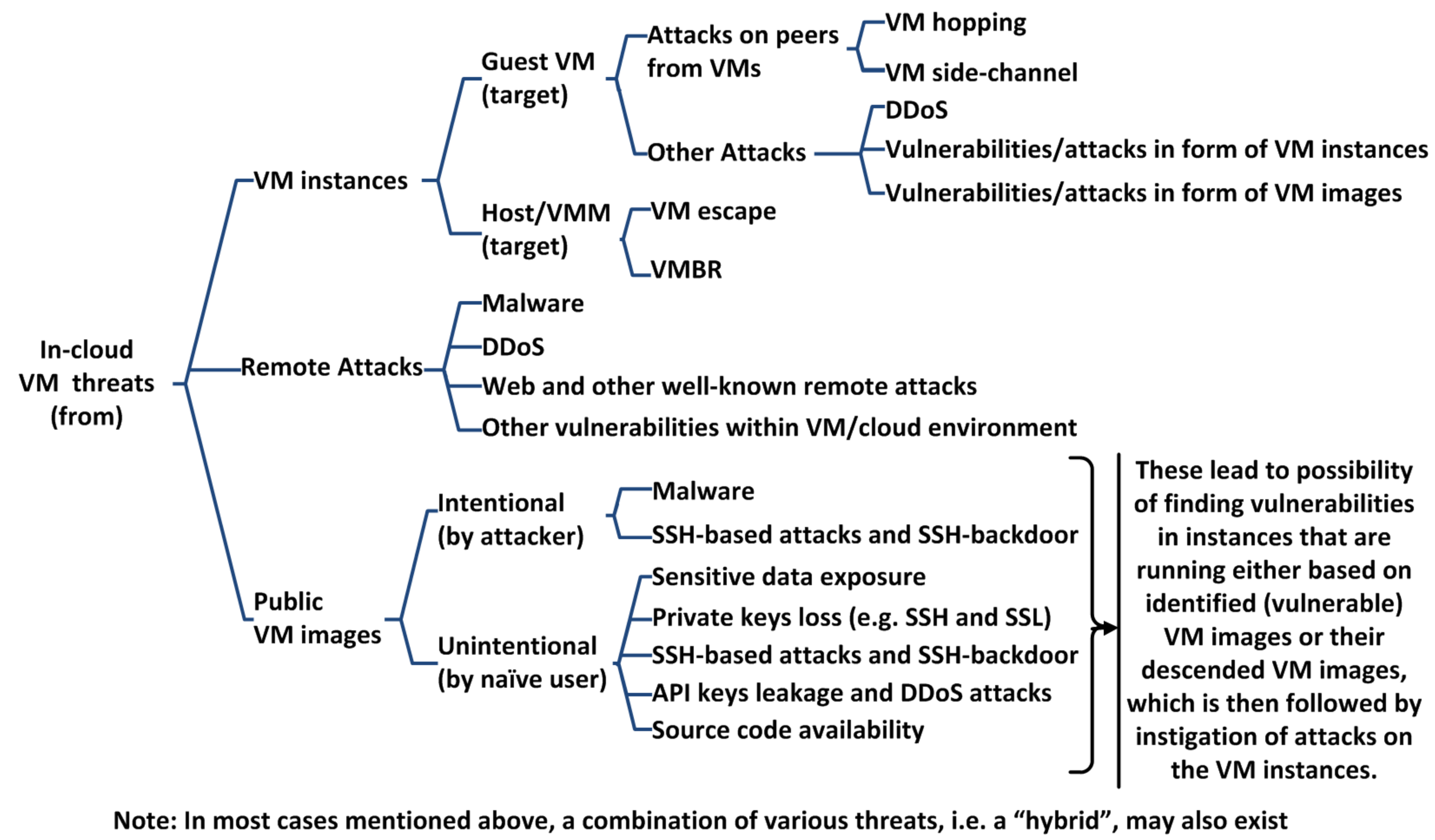}
\caption{Classification of threats to VMs in clouds based on the source of the threat. \tablename~\ref{tab:TableRelatedWorks} lists those works that have been surveyed in this article and which have also been used as a guide for the classification. Of particular note, is work presented by Bugiel et al. \cite{2011_Bugiel_VMimages}, whose classification comprising of four types of private-data loses owing to unintentional causes have been included in our classification above. Their work also discusses about SSH attacks (malicious or otherwise).}
\label{fig:ClassificationCloudVMAttacks}
\end{figure*}

On the basis of the survey presented in this article, a classification of VM threats in clouds has been shown in \figurename~\ref{fig:ClassificationCloudVMAttacks}. These have been divided into three main categories based on the source of the threat, as follows: 
\begin{inparaenum}[i)]
\item those security concerns occurring through live VM instances,
\item those which are raised due to remote attacks on VMs, and
\item those which are present within and propagate via publicly shared VM images.
\end{inparaenum}

The attacks (due to threats) instigated from VMs, may be directed toward guest VMs or their hypervisor and hosts. Also, the threats toward guest VM could further be categorized as per the eventual target and threat type, as follows:
\begin{inparaenum}[i)]
\item attacks on peer VMs that reside on the same physical machine (e.g. VM hopping and VM side-channel attacks), and
\item attacks on any VMs within a cloud environment due to VMs, (e.g. DDoS attacks from VMs due to API key leakage from VM images, other attacks from a live VM instances, or attacks caused by vulnerable/malicious live VMs being saved in the form of a publicly shared VM images, which are later moved, copied, used and/or instantiated).
\end{inparaenum}

Threats due to directed remote attacks on VMs through software vulnerabilities within VM, web attacks on VM, etc. have been discussed in Section~\ref{subsec:VMremote}. In addition, any general threats due to vulnerabilities within the cloud and VM environment (known or unknown at present) would fall within this category and exploited through remote attacks.

As for threats that begin from within VM images, they may be classified based on intentionality involved in creation of the threats, regarding which an in-depth discussion has already been presented in Section~\ref{subsec:VMimgesClouds}. The classification by Bugiel et al. \cite{2011_Bugiel_VMimages} has been the main source for this (VM images) part of the categorization, which we have shown pictorially alongside with categorization of threats from remote attacks, and live VM instance attacks. Regarding VM image threats, these may further be pushed to threats on other VM images and instances that are based on the originating culprit VM image.

From above discussion, hybrid threats may also emerge and the following gives an example one such instance:
\begin{inparaenum}[i)]
\item attacker gains access to a VM through a remote attack,
\item then uses VM hopping to compromise a peer VM,
\item followed by use of information resident within peer VM, i.e. data which was placed by the VM owner for personal use, to then gain access to users API keys, and
\item executing of DDoS on other systems using the API keys.
\end{inparaenum}
A related case would be an attacker directly obtaining API keys from a VM image and executing the DDoS attacks. These scenarios shows how ``recursive VM attacks'' or ``nested VM attacks'' may function, where one attack would enable another, followed by more, and once compromised, a VM may remain compromised until shutdown or other event that detects and blocks the attack, although is such latter cases confidential information is already lost to the attacker.

\section{Summary of surveyed works}
\label{sec:SummaryRelatedWorks}
\tablename~\ref{tab:TableRelatedWorks} provides a short summary of the publications that have been surveyed in this article and which have also been used as a guide for the classification.

\begin{longtable}{|l|l|p{0.575\textwidth}|}
\caption{List of virtualization security related works (in no particular order) }{\label{tab:TableRelatedWorks}} \\
\hline
\textbf{Year} & \textbf{Reference} & \textbf{Key notes} \\
\hline
\endfirsthead

\hline
\textbf{Year} & \textbf{Reference} & \textbf{Key notes} \\
\hline
\endhead

\multicolumn{3}{r}{\textit{Continued on next page}} \\
\endfoot
\hline
\endlastfoot

2006 & King et al. \cite{2006_SubVirt} & Introduces SubVirt. \\
\hline
2006 & Rutkowska \cite{2006_SuvertingVistaJR} &  Presents Blue Pill \cite{2006_IntrodBluePill} to subvert Windows Vista on AMD64. Lists differences between SubVirt and Blue Pill.\\
\hline
2006 & Zovi \cite{2006_Vitriol} & Presents Vitriol to subvert MacOS X on Intel Core Duo/Solo \\
\hline
2007 & Carpenter et al. \cite{2007_HidingVirtualizationFromAttackersMalware} & Briefly mentions the benefits of Virtualization. Discusses about VM espace attack. Two techniques for detection of VM (in VMware) are presented: 
\begin{inparaenum}[i)]
\item through VMware communications channel (e.g. Jerry.c tool), and
\item through Red Pill \cite{2007_VirtualisationAsBlackhatTool} tool (introduced by Joanna Rutkowska).
\end{inparaenum} 
Authors introduce VMmutate to prevent VM detection. \\
\hline
2007 & Skapinetz \cite{2007_VirtualisationAsBlackhatTool} & Discusses about Red Pill and Blue Pill \cite{2006_IntrodBluePill}.\\
\hline
2007 & Garfinkel et al. \cite{2007_CompatibilityIsNotTransparency} & Presents VMM Detection techniques.\\
\hline
2007 & Ford \& Allen \cite{2007_HowNotToBeSeenI}  & Classifies stealthy mechanisms used by malware: Passive, Active. Blue Pill and SubVirt are mentioned. \\
\hline
2007 & Allen \& Ford \cite{2007_HowNotToBeSeen-II}  & Part-II of Ford \& Allen \cite{2007_HowNotToBeSeenI}. Mechanisms to detect stealthy software by using ``cross-view diffs'' \cite{2007_HowNotToBeSeen-II} technique and lists software that use uses the technique. Highlights techniques mentioned by Garfinkel et al. \cite{2007_CompatibilityIsNotTransparency} for detection of Hypervisors thereby attacks such as SubVirt and Blue Pill could be circumvented.\\
\hline
2008 & Carbone et al. \cite{2008_TamingVirtualization}  & Discusses about VMBRs: SubVirt, Vitriol, Blue Pill. Actions by VMBRs are termed as ``hyperjacking'' \cite{2008_TamingVirtualization}. Proposes GuardHype as a concept to prevent VMBR attacks by leveraging VM control structures (VMCS) in a nested hypervisor setup. \\
\hline
2008 & Nance et al. \cite{2008_VirtualMachineIntrospectionObsInterference} & Work concerns itself with research in virtual machine introspection (VMI): a countermeasure technique that keeps an eye on activities within VM.  \\
\hline
2009 & Ristenpart et al. \cite{2009_HeyYouGetOffCloud} & Compelling work highlighting vulnerabilities in VM environments due to multitenancy on physical machines in public clouds, and building on the information to introduce ``cross-VM side-channel attacks'' \cite{2009_HeyYouGetOffCloud}. The authors orchestrate the attacks on Amazon EC2 cloud. The mentioned attacks include ``cross-VM keystroke monitoring'' \cite{2009_HeyYouGetOffCloud} and provide means of gathering confidential information. Few suggestions on how to prevent them are provided.\\
\hline
2009 & Christodorescu et al. \cite{2009_CloudSecurityIsNotJustVirtSecurity} & Introduces their approach of VM introspection that requires no prior knowledge of the guest OS running on the VMs or their security state, which was lacking in the previous introspection techniques. Their approach is useful specially in dynamic environments such as cloud computing where VM migration is common and security of the VM has to be evaluated, and managed. The authors two applications of their introspection technique: 
\begin{inparaenum}[i)]
\item VM OS identification, and
\item Rootkit detection and recovery.
\end{inparaenum} 
\\
\hline
2009 & Provos et al. \cite{2009_CyberCrimeWhenCloudTurnsDark} & This article focuses on cyber-crimes. The shift in motivation of attackers, attacks on web-browsers and web servers are discussed. Discussions include: ``drive-by download'' \cite{2009_CyberCrimeWhenCloudTurnsDark}, SQL injection attacks, .htaccess redirection attacks, social engineering attacks, and challenges. Although this article doesn't mention about cloud computing, the issues highlighted are quite relevant in the cloud domain and should not be ignored. \\
\hline
2010 & Jasti et al. \cite{2010_SecurityInMultiTenancyCloud} & This paper focuses toward security due to cloud multitenancy. Attacks and vulnerabilities such as VM hopping, VM escape and VM mobility are discussed. The authors setup a Eucalyptus cloud with Xen hypervisor test-bed and verify if VMs are allocated fair share of network, CPU and memory resources. \\
\hline
2010 & Anthes \cite{2010_SecurityInTheCloud}  & Discusses about virtual machine introspection\\
\hline
2010 & Bradbury \cite{2010_VirtualAttacks}  &  Focuses on virtualization security issues related to cloud computing. Blue Pill attack is discussed here.\\
\hline
2011 & Bugiel et al. \cite{2011_Bugiel_VMimages} & A detailed study about how publicly shared VM images lead to security threats in the cloud environment, discussing how about unintentional actions by na\"{\i}ve users can cause various issues. Malicious causes for the threats have also been mentioned along with discussion of related works regarding VM image security.\\
\hline
2012 & Tsai et al. \cite{2012_ThreatAsAServiceVirtImpact} & Discusses about VM hopping, VM mobility and other vulnerabilities, including their impact on confidentiality, integrity and availability at different cloud service levels.\\
\hline
2012 & McClure et al. \cite{2012_mcclure_hackingExposed7} & A comprehensive guide explaining from ground up about how attackers compromise different types of systems by employing a variety of techniques and remote attacks and at the same time discussing the tools that are employed to make the attacking jobs simple. For security of users, safeguards against attacks have been discussed as well. \\
\end{longtable}

\section{Observations on cloud virtualization security}
\label{sec:ObsCloudSecurity}
From the survey above, a pattern emerges as to how security in cloud virtualization has been evolving. Research in VM detection, VM detection avoidance, advances in VMBR, followed by attempts on detection of VMBR, appear to follow a loop, each technique trying to evade or stay on top of the other. Malwares are trying to hide from being identified and VMs are hiding from being detected, both making an effort to dodge each other. Carpenter et al. \cite{2007_HidingVirtualizationFromAttackersMalware} with regards to VM detection discuss how it can develop into ``a dangerous game of cat and mouse \ldots'' \cite{2007_HidingVirtualizationFromAttackersMalware}. On a similar note, King et al. \cite{2006_SubVirt} while introducing the topic of VMBRs mention about a ``battle \ldots\ taking place between attackers and defenders \ldots'' \cite{2006_SubVirt}, discussing the nature of attackers, who are trying to hide from defenders, and the latter in turn being actively in pursuit of attackers footprints. These remarks are pretty fair assessment of the game that is in action. Interestingly, both the attackers and defenders are motivating each other to perform their best.

On a similar note as above, security researchers are playing the role of pseudo-attackers.  King et al. \cite{2006_SubVirt}, Rutkowska \cite{2012_InvisibleThingsLab,2012_InvisibleBlogSpot,2006_IntrodBluePill,2006_BlogLog}, Zovi \cite{2006_Vitriol}, Ristenpart et al. \cite{2009_HeyYouGetOffCloud} and Bugiel et al. \cite{2011_Bugiel_VMimages} have all contributed significantly toward improving of virtualization security --- thereby indirectly helping cloud security --- by first highlighting the vulnerabilities, then exploiting those vulnerabilities through novel attacks, subsequently suggesting approaches to mitigate those attacks and providing future research directions to further improve the system security. As long as researchers stay ahead of the real malicious attackers, cloud security is under check.

VM image based attacks demonstrate how easily vulnerabilities could be introduced and spread throughout the cloud system, also highlighting how relevant it is to safeguard against such threats. Furthermore, they also demonstrate how vulnerabilities still remain effective in different states of VMs  --- i.e. in two states: VM instance and VM image --- and remain even after transformation from one state to the other, e.g. vulnerabilities that are present in a running VM instance would still remain a threat if VM instance is transformed and saved as a VM image. However, the threat would be lot greater and amplified, if lot many instances were to then derive from the same vulnerable/malicious VM image.

With regards to VMs, it can be readily realized that the barriers they provide are nothing but virtual. These barriers may be broken and attacks such as VM escape and VM hopping clearly demonstrate such weaknesses. Measures should be in place to enforce these barriers, such that any action to circumvent them are caught. Further advances and research in this area is needed.

An important point to note is that although attacks such as VM escape, VM hopping, etc. remain as relevant in private clouds as in public clouds, there appears to be a difference is the risk probability. Public clouds allow access to customers all over the globe, whereas private clouds are restrictive to an organization, running in their local cloud network. The malicious attacks from unknown malicious customers is not a concern in private clouds. However, the threat in private clouds is more focused toward insider threats \cite{2010_CSATopThreats}. An organization considering migration to the clouds may consider other possibilities instead of just public clouds, such as private or hybrid clouds. Research into differences between risks related to private, hybrid and public clouds and their in-depth comparison would help in this regard.

Lastly, it is key to see clouds as an achievement and have a ``positive''\cite{2011_PositiveCloudSecurity} outlook toward it. After all, cloud computing is the realization of a foresight about sharing computing as a ``utility'' \cite{2012_JohnMcCarthy,2009_InCloudsShallWeTrust,2010_SecurityInTheCloud,zhang2010cloud13}.

\section{Conclusions}
\label{sec:Conclusions}
In this article, virtualization security threats in the cloud computing domain were reviewed. From previous publications, we considered attacks and vulnerabilities encountered in virtual environments and described how these attacks related to cloud systems. In addition, distinctive cloud virtualization threats were also highlighted. The focus has been on virtual machine (VM) security where attacks and vulnerabilities such as VM escape, VM hopping, VM side-channel, VM-based rootkits (VMBRs), VM mobility, VM remote and cyber attacks were discussed. Of the works and the attacks surveyed, the following are the most cloud-specific threats: ``VM hopping'' \cite{2010_SecurityInMultiTenancyCloud,2012_ThreatAsAServiceVirtImpact}, ``cross-VM side-channel attacks''\cite{2009_HeyYouGetOffCloud}, ``VM Mobility'' \cite{2012_ThreatAsAServiceVirtImpact}, and other VM image attacks \cite{2011_Bugiel_VMimages}. Countermeasures and security measures to detect or prevent these attacks through techniques such as VM detection, GuardHype, VM introspection, VM image scanning, etc. were discussed. On the basis of the survey presented in this article, a classification of VM threats in clouds was presented. Finally, we included our own views and those of other researchers on the topic.

\bibliographystyle{elsarticle-num}
\bibliography{computerSecurity}

\begin{thebibliography}{10}
\expandafter\ifx\csname url\endcsname\relax
  \def\url#1{\texttt{#1}}\fi
\expandafter\ifx\csname urlprefix\endcsname\relax\def\urlprefix{URL }\fi
\expandafter\ifx\csname href\endcsname\relax
  \def\href#1#2{#2} \def\path#1{#1}\fi

\bibitem{2009_CloutCloud}
J.~Hayes, Clout of the cloud, Engineering Technology 4~(6) (2009) 60--61.

\bibitem{2009_CloudNewWineNewBottle}
J.~Voas, J.~Zhang, Cloud computing: New wine or just a new bottle?, IT
  Professional 11~(2) (2009) 15--17.

\bibitem{2010_SecuringElasticityCloud}
D.~Owens, \href{http://doi.acm.org/10.1145/1743546.1743565}{Securing elasticity
  in the cloud}, Commun. ACM 53~(6) (2010) 46--51.
\newblock \href {http://dx.doi.org/10.1145/1743546.1743565}
  {\path{doi:10.1145/1743546.1743565}}.
\newline\urlprefix\url{http://doi.acm.org/10.1145/1743546.1743565}

\bibitem{2010_OverviewVirtualCloudComputing}
H.~F. Cervone, An overview of virtual and cloud computing, OCLC Systems and
  Services 26~(3) (2010) 162--165.

\bibitem{zhang2010cloud13}
Q.~Zhang, L.~Cheng, R.~Boutaba, Cloud computing: state-of-the-art and research
  challenges, Journal of Internet Services and Applications 1~(1) (2010) 7--18.

\bibitem{2008_GridComputingSecurityTaxonomy}
A.~Chakrabarti, A.~Damodaran, S.~Sengupta, Grid computing security: A taxonomy,
  Security Privacy, IEEE 6~(1) (2008) 44--51.
\newblock \href {http://dx.doi.org/10.1109/MSP.2008.12}
  {\path{doi:10.1109/MSP.2008.12}}.

\bibitem{2012_JohnMcCarthy}
R.~Seising,
  \href{http://www.sciencedirect.com/science/article/pii/S0933365712000164}{{John
  McCarthy, 1927-2011}}, Artificial Intelligence in Medicine 54~(3) (2012)
  151--154.
\newblock \href {http://dx.doi.org/10.1016/j.artmed.2012.01.003}
  {\path{doi:10.1016/j.artmed.2012.01.003}}.
\newline\urlprefix\url{http://www.sciencedirect.com/science/article/pii/S0933365712000164}

\bibitem{2009_InCloudsShallWeTrust}
B.~Michael, \href{http://dx.doi.org/10.1109/MSP.2009.124}{In clouds shall we
  trust?}, IEEE Security and Privacy 7~(5) (2009) 3.
\newblock \href {http://dx.doi.org/10.1109/MSP.2009.124}
  {\path{doi:10.1109/MSP.2009.124}}.
\newline\urlprefix\url{http://dx.doi.org/10.1109/MSP.2009.124}

\bibitem{2010_SecurityInTheCloud}
G.~Anthes, \href{http://doi.acm.org/10.1145/1839676.1839683}{Security in the
  cloud}, Commun. ACM 53~(11) (2010) 16--18.
\newblock \href {http://dx.doi.org/10.1145/1839676.1839683}
  {\path{doi:10.1145/1839676.1839683}}.
\newline\urlprefix\url{http://doi.acm.org/10.1145/1839676.1839683}

\bibitem{2012_ucc2012YasDas}
Y.~Shoaib, O.~Das,
  \href{http://ieeexplore.ieee.org/stamp/stamp.jsp?tp=&arnumber=6424935&isnumber=6424908}{Using
  layered bottlenecks for virtual machine provisioning in the clouds}, in:
  Proceedings of the 2012 IEEE/ACM Fifth International Conference on Utility
  and Cloud Computing (UCC 2012), 2012, pp. 109--116.
\newblock \href {http://dx.doi.org/10.1109/UCC.2012.10}
  {\path{doi:10.1109/UCC.2012.10}}.
\newline\urlprefix\url{http://ieeexplore.ieee.org/stamp/stamp.jsp?tp=&arnumber=6424935&isnumber=6424908}

\bibitem{NIST_CloudDefn}
P.~Mell, T.~Grance,
  \href{http://csrc.nist.gov/publications/nistpubs/800-145/SP800-145.pdf}{The
  {NIST} definition of cloud computing}, Special Publication 800-145, NIST
  (National Institute of Standards and Technology), Gaithersburg, MD
  20899-8930, [Accessed March 14, 2013] (Sep 2011).
\newline\urlprefix\url{http://csrc.nist.gov/publications/nistpubs/800-145/SP800-145.pdf}

\bibitem{2012_MeasureElasticity}
S.~Islam, K.~Lee, A.~Fekete, A.~Liu,
  \href{http://doi.acm.org/10.1145/2188286.2188301}{How a consumer can measure
  elasticity for cloud platforms}, in: Proceedings of the third joint
  WOSP/SIPEW international conference on Performance Engineering, ICPE '12,
  ACM, New York, NY, USA, 2012, pp. 85--96.
\newblock \href {http://dx.doi.org/10.1145/2188286.2188301}
  {\path{doi:10.1145/2188286.2188301}}.
\newline\urlprefix\url{http://doi.acm.org/10.1145/2188286.2188301}

\bibitem{2013_GoogleTrends}
Google, \href{http://www.google.ca/trends/}{Google trends}, [Accessed Feb. 12,
  2013]. \copyright 2013 Google. Reprinted, with permission, from Google
  (2013).
\newline\urlprefix\url{http://www.google.ca/trends/}

\bibitem{2009_CloudSecurityInfoWeek}
P.~Korzeniowski, M.~Jander, Cloud security, InformationWeek~(1247) (2009)
  HB4--HB6,HB8,HB10,HB12,HB14.

\bibitem{2012_SecurityChallengesPublicCloud}
K.~Ren, C.~Wang, Q.~Wang, Security challenges for the public cloud, Internet
  Computing, IEEE 16~(1) (2012) 69--73.
\newblock \href {http://dx.doi.org/10.1109/MIC.2012.14}
  {\path{doi:10.1109/MIC.2012.14}}.

\bibitem{2010_TowarsMultiTenancyAuthCloud}
J.~Calero, N.~Edwards, J.~Kirschnick, L.~Wilcock, M.~Wray, Toward a
  multi-tenancy authorization system for cloud services, Security Privacy, IEEE
  8~(6) (2010) 48--55.
\newblock \href {http://dx.doi.org/10.1109/MSP.2010.194}
  {\path{doi:10.1109/MSP.2010.194}}.

\bibitem{2012_ThreatAsAServiceVirtImpact}
H.-Y. Tsai, M.~Siebenhaar, A.~Miede, Y.~Huang, R.~Steinmetz, Threat as a
  service?: Virtualization's impact on cloud security, IT Professional 14~(1)
  (2012) 32--37.
\newblock \href {http://dx.doi.org/10.1109/MITP.2011.117}
  {\path{doi:10.1109/MITP.2011.117}}.

\bibitem{2012_EnforcingMultitenancyCloud}
J.~Fiaidhi, I.~Bojanova, J.~Zhang, L.-J. Zhang, Enforcing multitenancy for
  cloud computing environments, IT Professional 14~(1) (2012) 16--18.
\newblock \href {http://dx.doi.org/10.1109/MITP.2012.6}
  {\path{doi:10.1109/MITP.2012.6}}.

\bibitem{ranjan2010peer12}
R.~Ranjan, L.~Zhao, X.~Wu, A.~Liu, A.~Quiroz, M.~Parashar, Peer-to-peer cloud
  provisioning: Service discovery and load-balancing, Cloud Computing (2010)
  195--217.

\bibitem{velte2009cloud1}
T.~Velte, A.~Velte, R.~Elsenpeter, Cloud computing: a practical approach,
  McGraw-Hill Osborne Media, 2009.

\bibitem{2005_ResourceVirtRenaissance}
R.~Figueiredo, P.~Dinda, J.~Fortes, Guest editors' introduction: Resource
  virtualization renaissance, Computer 38~(5) (2005) 28--31.
\newblock \href {http://dx.doi.org/10.1109/MC.2005.159}
  {\path{doi:10.1109/MC.2005.159}}.

\bibitem{2005_IntelVT}
R.~Uhlig, G.~Neiger, D.~Rodgers, A.~Santoni, F.~Martins, A.~Anderson,
  S.~Bennett, A.~Kagi, F.~Leung, L.~Smith, Intel virtualization technology,
  Computer 38~(5) (2005) 48--56.
\newblock \href {http://dx.doi.org/10.1109/MC.2005.163}
  {\path{doi:10.1109/MC.2005.163}}.

\bibitem{2005_VirtualMachineMonitorsCurrentTechFutureTrends}
M.~Rosenblum, T.~Garfinkel, Virtual machine monitors: current technology and
  future trends, Computer 38~(5) (2005) 39--47.
\newblock \href {http://dx.doi.org/10.1109/MC.2005.176}
  {\path{doi:10.1109/MC.2005.176}}.

\bibitem{2010_InVMMeasuringFramework}
Q.~Liu, C.~Weng, M.~Li, Y.~Luo, An in-vm measuring framework for increasing
  virtual machine security in clouds, Security Privacy, IEEE 8~(6) (2010)
  56--62.
\newblock \href {http://dx.doi.org/10.1109/MSP.2010.143}
  {\path{doi:10.1109/MSP.2010.143}}.

\bibitem{2008_MeetTheVirts}
T.~Killalea, \href{http://doi.acm.org/10.1145/1348583.1348589}{Meet the virts},
  Queue 6~(1) (2008) 14--18.
\newblock \href {http://dx.doi.org/10.1145/1348583.1348589}
  {\path{doi:10.1145/1348583.1348589}}.
\newline\urlprefix\url{http://doi.acm.org/10.1145/1348583.1348589}

\bibitem{2008_ParadoxOfSecurityInVM}
M.~Price, The paradox of security in virtual environments, Computer 41~(11)
  (2008) 22--28.
\newblock \href {http://dx.doi.org/10.1109/MC.2008.472}
  {\path{doi:10.1109/MC.2008.472}}.

\bibitem{2010_Sempolinski_ComparisonOpenEucNim}
P.~Sempolinski, D.~Thain, A comparison and critique of eucalyptus, opennebula
  and nimbus, in: Proceedings of the 2010 IEEE Second International Conference
  on Cloud Computing Technology and Science, CLOUDCOM '10, IEEE Computer
  Society, Washington, DC, USA, 2010, pp. 417--426.

\bibitem{2013_AmazonAMI}
{Amazon Web Services}, \href{https://aws.amazon.com/amis/}{{Amazon Machine
  Images}}, [Accessed Aug 2, 2013] (2013).
\newline\urlprefix\url{https://aws.amazon.com/amis/}

\bibitem{2013_AmazonEC2}
{Amazon Web Services}, \href{http://aws.amazon.com/ec2/}{{Amazon Elastic
  Compute Cloud (Amazon EC2)}}, [Accessed Aug 4, 2013] (2013).
\newline\urlprefix\url{http://aws.amazon.com/ec2/}

\bibitem{2011_Bugiel_VMimages}
S.~Bugiel, S.~N\"{u}rnberger, T.~P\"{o}ppelmann, A.-R. Sadeghi, T.~Schneider,
  Amazonia: when elasticity snaps back, in: Proceedings of the 18th ACM
  conference on Computer and communications security, CCS '11, 2011, pp.
  389--400.

\bibitem{2008_TamingVirtualization}
M.~Carbone, W.~Lee, D.~Zamboni, Taming virtualization, Security Privacy, IEEE
  6~(1) (2008) 65--67.
\newblock \href {http://dx.doi.org/10.1109/MSP.2008.24}
  {\path{doi:10.1109/MSP.2008.24}}.

\bibitem{2008_VirtSparksSecurityConcerns}
S.~Vaughan-Nichols, Virtualization sparks security concerns, Computer 41~(8)
  (2008) 13--15.
\newblock \href {http://dx.doi.org/10.1109/MC.2008.276}
  {\path{doi:10.1109/MC.2008.276}}.

\bibitem{2007_HidingVirtualizationFromAttackersMalware}
M.~Carpenter, T.~Liston, E.~Skoudis, Hiding virtualization from attackers and
  malware, Security Privacy, IEEE 5~(3) (2007) 62--65.
\newblock \href {http://dx.doi.org/10.1109/MSP.2007.63}
  {\path{doi:10.1109/MSP.2007.63}}.

\bibitem{2009_HeyYouGetOffCloud}
T.~Ristenpart, E.~Tromer, H.~Shacham, S.~Savage,
  \href{http://doi.acm.org/10.1145/1653662.1653687}{Hey, you, get off of my
  cloud: exploring information leakage in third-party compute clouds}, in:
  Proceedings of the 16th ACM conference on Computer and communications
  security, CCS '09, ACM, New York, NY, USA, 2009, pp. 199--212.
\newblock \href {http://dx.doi.org/10.1145/1653662.1653687}
  {\path{doi:10.1145/1653662.1653687}}.
\newline\urlprefix\url{http://doi.acm.org/10.1145/1653662.1653687}

\bibitem{2010_SecurityInMultiTenancyCloud}
A.~Jasti, P.~Shah, R.~Nagaraj, R.~Pendse, Security in multi-tenancy cloud, in:
  Security Technology (ICCST), 2010 IEEE International Carnahan Conference on,
  2010, pp. 35--41.
\newblock \href {http://dx.doi.org/10.1109/CCST.2010.5678682}
  {\path{doi:10.1109/CCST.2010.5678682}}.

\bibitem{2006_IntrodBluePill}
J.~Rutkowska,
  \href{http://theinvisiblethings.blogspot.ca/2006/06/introducing-blue-pill.html}{{Introducing
  Blue Pill}}, [Accessed March 14, 2013] (June 2006).
\newline\urlprefix\url{http://theinvisiblethings.blogspot.ca/2006/06/introducing-blue-pill.html}

\bibitem{2006_SuvertingVistaJR}
J.~Rutkowska,
  \href{http://www.blackhat.com/presentations/bh-jp-06/BH-JP-06-Rutkowska.pdf}{{Subverting
  Vista\texttrademark\ Kernel For Fun And Profit}}, {Black Hat Japan 2006
  (Presentation)}, [Accessed March 14, 2013] (Oct 5 2006).
\newline\urlprefix\url{http://www.blackhat.com/presentations/bh-jp-06/BH-JP-06-Rutkowska.pdf}

\bibitem{2010_CSATopThreats}
{CSA (Cloud Security Alliance)},
  \href{https://cloudsecurityalliance.org/topthreats/csathreats.v1.0.pdf}{Top
  threats to cloud computing v1.0}, [Accessed March 14, 2013] (March 2010).
\newline\urlprefix\url{https://cloudsecurityalliance.org/topthreats/csathreats.v1.0.pdf}

\bibitem{2007_VirtualisationAsBlackhatTool}
K.~Skapinetz, Virtualisation as a blackhat tool, Network Security 2007~(10)
  (2007) 4--7.

\bibitem{2012_InvisibleThingsLab}
{Invisible Things Lab (ITL)},
  \href{http://invisiblethingslab.com/itl/About.html}{About the company},
  [Accessed March 14, 2013] (2007-2012).
\newline\urlprefix\url{http://invisiblethingslab.com/itl/About.html}

\bibitem{2012_InvisibleBlogSpot}
J.~Rutkowska, \href{http://theinvisiblethings.blogspot.ca}{{The Invisible
  Things Lab's blog}}, [Accessed March 14, 2013] (2006-2013).
\newline\urlprefix\url{http://theinvisiblethings.blogspot.ca}

\bibitem{2010_VirtualizationSurveyConceptsTaxonomy}
J.~Sahoo, S.~Mohapatra, R.~Lath, Virtualization: A survey on concepts, taxonomy
  and associated security issues, in: Computer and Network Technology (ICCNT),
  2010 Second International Conference on, 2010, pp. 222--226.
\newblock \href {http://dx.doi.org/10.1109/ICCNT.2010.49}
  {\path{doi:10.1109/ICCNT.2010.49}}.

\bibitem{2007_HowNotToBeSeenI}
R.~Ford, W.~H. Allen, {How Not to Be Seen}, Security Privacy, IEEE 5~(1) (2007)
  67--69.
\newblock \href {http://dx.doi.org/10.1109/MSP.2007.8}
  {\path{doi:10.1109/MSP.2007.8}}.

\bibitem{2007_HowNotToBeSeen-II}
W.~Allen, R.~Ford, {How Not to Be Seen II: The Defenders Fight Back}, Security
  Privacy, IEEE 5~(6) (2007) 65--68.
\newblock \href {http://dx.doi.org/10.1109/MSP.2007.166}
  {\path{doi:10.1109/MSP.2007.166}}.

\bibitem{2006_SubVirt}
S.~T. King, P.~M. Chen, Y.-M. Wang, C.~Verbowski, H.~J. Wang, J.~R. Lorch,
  \href{http://dx.doi.org/10.1109/SP.2006.38}{{SubVirt: Implementing malware
  with virtual machines}}, in: Proceedings of the 2006 IEEE Symposium on
  Security and Privacy, SP '06, IEEE Computer Society, Washington, DC, USA,
  2006, pp. 314--327.
\newblock \href {http://dx.doi.org/10.1109/SP.2006.38}
  {\path{doi:10.1109/SP.2006.38}}.
\newline\urlprefix\url{http://dx.doi.org/10.1109/SP.2006.38}

\bibitem{2006_Vitriol}
D.~D. Zovi,
  \href{http://www.blackhat.com/presentations/bh-usa-06/BH-US-06-Zovi.pdf}{{Hardware
  Virtualization Rootkits}}, {Black Hat USA 2006 Briefing \& Trainings
  (Presentation)}, [Accessed March 14, 2013] (Jul 29-Aug 3 2006).
\newline\urlprefix\url{http://www.blackhat.com/presentations/bh-usa-06/BH-US-06-Zovi.pdf}

\bibitem{2006_BlogLog}
J.~Rutkowska, Blog log, Network Computing 17~(14) (2006) 16.

\bibitem{2008_KickHypervisor}
J.~Rutkowska,
  \href{http://theinvisiblethings.blogspot.ca/2008/03/kick-ass-hypervisor-nesting.html}{{Kick
  Ass Hypervisor Nesting!}}, [Accessed March 14, 2013] (March 2008).
\newline\urlprefix\url{http://theinvisiblethings.blogspot.ca/2008/03/kick-ass-hypervisor-nesting.html}

\bibitem{2010_VirtualAttacks}
D.~Bradbury, Virtual attacks, Infosecurity 7~(6) (2010) 32--35.

\bibitem{2008_Vegas2008}
J.~Rutkowska,
  \href{http://theinvisiblethings.blogspot.ca/2008/04/vegas-training-2008.html}{{Vegas
  Training 2008}}, [Accessed March 14, 2013] (April 2008).
\newline\urlprefix\url{http://theinvisiblethings.blogspot.ca/2008/04/vegas-training-2008.html}

\bibitem{2007_CompatibilityIsNotTransparency}
T.~Garfinkel, K.~Adams, A.~Warfield, J.~Franklin,
  \href{http://static.usenix.org/event/hotos07/tech/full_papers/garfinkel/garfinkel.pdf}{{Compatibility
  is not transparency: VMM detection myths and realities}}, in: Proceedings of
  the 11th USENIX workshop on Hot topics in operating systems, HOTOS'07, USENIX
  Association, Berkeley, CA, USA, 2007, pp. 6:1--6:6.
\newline\urlprefix\url{http://static.usenix.org/event/hotos07/tech/full_papers/garfinkel/garfinkel.pdf}

\bibitem{2012_mcclure_hackingExposed7}
S.~McClure, J.~Scambray, G.~Kurtz, Hacking Exposed 7 : Network Security Secrets
  \& Solutions, 7th Edition, Hacking Exposed, McGraw-Hill, Inc., New York, NY,
  USA, 2012.

\bibitem{2009_CyberCrimeWhenCloudTurnsDark}
N.~Provos, M.~A. Rajab, P.~Mavrommatis,
  \href{http://doi.acm.org/10.1145/1498765.1498782}{Cybercrime 2.0: when the
  cloud turns dark}, Commun. ACM 52~(4) (2009) 42--47.
\newblock \href {http://dx.doi.org/10.1145/1498765.1498782}
  {\path{doi:10.1145/1498765.1498782}}.
\newline\urlprefix\url{http://doi.acm.org/10.1145/1498765.1498782}

\bibitem{2008_VirtualMachineIntrospectionObsInterference}
K.~Nance, M.~Bishop, B.~Hay, Virtual machine introspection: Observation or
  interference?, Security Privacy, IEEE 6~(5) (2008) 32--37.
\newblock \href {http://dx.doi.org/10.1109/MSP.2008.134}
  {\path{doi:10.1109/MSP.2008.134}}.

\bibitem{2009_CloudSecurityIsNotJustVirtSecurity}
M.~Christodorescu, R.~Sailer, D.~L. Schales, D.~Sgandurra, D.~Zamboni,
  \href{http://doi.acm.org/10.1145/1655008.1655022}{Cloud security is not
  (just) virtualization security: a short paper}, in: Proceedings of the 2009
  ACM workshop on Cloud computing security, CCSW '09, ACM, New York, NY, USA,
  2009, pp. 97--102.
\newblock \href {http://dx.doi.org/10.1145/1655008.1655022}
  {\path{doi:10.1145/1655008.1655022}}.
\newline\urlprefix\url{http://doi.acm.org/10.1145/1655008.1655022}

\bibitem{2011_PositiveCloudSecurity}
P.~Wilson,
  \href{http://www.sciencedirect.com/science/article/pii/S1363412711000471}{Positive
  perspectives on cloud security}, Information Security Technical Report
  16~(3–4) (2011) 97--101, (Cloud Security).
\newblock \href {http://dx.doi.org/10.1016/j.istr.2011.08.002}
  {\path{doi:10.1016/j.istr.2011.08.002}}.
\newline\urlprefix\url{http://www.sciencedirect.com/science/article/pii/S1363412711000471}

\end{thebibliography}

\end{document}